\documentstyle[12pt]{article}

\begin{document}

\thispagestyle{empty} \renewcommand{\thefootnote}{\fnsymbol{footnote}}

\begin{center}
\vspace{1cm}{\Large {\bf TOWARDS SO(2,10)-INVARIANT $M$-THEORY:
MULTILAGRANGIAN FIELDS}}

\vspace{1cm} {\bf R. Manvelyan}$^{1}$ \footnote{%
E-mail: manvel@physik.uni-kl.de} \footnote{%
Alexander von Humboldt Fellow,
\par
On leave from Yerevan Physics Istitute,} and {\bf R. Mkrtchyan}$^{2}$
\footnote{%
E-mail: mrl@amsun.yerphi.am} \vspace{1cm}

$^{1}${\it Department of Physics, University of Kaiserslautern,}

{\it P.\ O.\ Box 3049, D 67653 Kaiserslautern, Germany} \vspace{1cm}

$^{2}${\it Theoretical Physics Department,} {\it Yerevan Physics Institute,}

{\it Alikhanian Br. St.2, Yerevan, 375036 Armenia }
\end{center}

\vspace{1cm}
\begin{abstract}
The $SO(2,10)$ covariant extension of M-theory superalgebra is considered,
with the aim to construct a correspondingly generalized $M$-theory, or $11d$
supergravity. For the orbit, corresponding to the $11d$ supergravity
multiplet, the simplest unitary representations of the bosonic part of this
algebra, with sixth-rank tensor excluded, are constructed on a language of
field theory in $66d$ space-time. The main peculiarities are the presence of
more than one equation of motion and corresponding Lagrangians for a given
field and that the gauge and SUSY invariances of the theory mean that the
sum of variations of these Lagrangians (with different variations of the
same field) is equal to zero.
\end{abstract}
\renewcommand{\thefootnote}{\arabic{footnote}} \setcounter{footnote}0

{\smallskip \pagebreak }

\section{Introduction}

Recent progress in the investigation of theories with maximal supersymmetry
leads to the notions of $M$ and $F$ theories\cite{F}, unifying many
previously disconnected ones, such as $11d$ supergravity, superstrings,
branes, etc. One of the lessons of this progress is that all features of the
supersymmetry algebra appear to be important and contain interesting
information on the physics behind them. Here we make an attempt to develop a
systematic approach, aimed at using the well known property of maximally
extended $11d$ supersymmetry algebra, namely its invariance with respect to
$%
SO(2,10)$ rotations \cite{B}. This feature has been discussed in a number of
papers \cite{MM} - \cite{Hef1}, where, in particular, twelve-dimensional
theories were constructed, the particles and branes models, etc. What is the
motivation for an $SO(2,10)$ approach to the $M$- theory superalgebra?

The $11$-dimensional supersymmetry algebra with maximal number of
``central'' charges \cite{Tow} (which are not central but rather are tensors
with respect to the space-time rotations) has the following anticommutator
of supercharges
\begin{eqnarray}
\left\{ \bar{Q},Q\right\} &=&\Gamma ^{i}P_{i}+\Gamma ^{ij}Z_{ij}+\Gamma
^{ijklm}Z_{ijklm},  \label{1} \\
i,j,... &=&0,1,2,..10  \nonumber
\end{eqnarray}
Here the $Q$'s are $11d$ Majorana spinors, $P_{i}$ are the usual momenta,
and tensors $Z_{ij},Z_{ijklm}$ the abovementioned central charges. The
interesting feature of this relation is its $SO(2,10)$ invariance, which
means that it is invariant with respect to the $SO(2,10)$ group of rotations
of space-time with two ``times'' and 10 space coordinates. Thus $11$
dimensional Majorana spinors $Q$ can be identified as $12d$ Majorana-Weyl
spinors, and generators on the r.h.s. combine into a $12d$ tensor $P_{\mu
\nu }$ and a selfdual sixth-rank tensor $Z_{\mu \nu \lambda \rho \sigma
\delta }^{+}$. Instead of $(\ref{1})$ we have now in 12d notation
\begin{eqnarray}
\left\{ \bar{Q},Q\right\} &=&\Gamma ^{\mu \nu }P_{\mu \nu }+\Gamma ^{\mu \nu
\lambda \rho \sigma \delta }Z_{\mu \nu \lambda \rho \sigma \delta }^{+}
\label{2} \\
\mu \nu ,... &=&0^{\prime },0,1,...10  \nonumber
\end{eqnarray}
Reducing the right hand side from $12d$ to $11d$ notation we obtain again
relation (\ref{1}). The important difference between these two relations is
the absence in the second one of the vector object, momentum $P_{i}$. We
ignore first supercharges $Q$ , and consider the pure ``Poincare'' algebra,
implied by $(\ref{2})$, which now consists of the usual generators of $%
SO(2,10)$ rotations and tensors $P_{\mu \nu }$ with evident commutation
relations: components of $P_{\mu \nu }$ are mutually commutative, and
transform as second rank tensors under rotations. We shall obtain features,
which differ considerably from the usual ones. The main difference originats
from the presence of many invariants (Casimir's operators), constructed from
$P_{\mu \nu}$:
\begin{eqnarray}
TrP^{2} &=&P_{\mu \nu }P^{\nu \mu },  \nonumber \\
TrP^{4} &=&P_{\mu \nu }P^{\nu \lambda }P_{\lambda \rho }P^{\rho \mu },
\label{3} \\
&...&  \nonumber \\
TrP^{12} &=&P_{\mu \mu _{1}}P^{\mu _{1}\mu _{2}}...P^{\mu _{11}\mu }.
\nonumber
\end{eqnarray}
instead of the single one in $11d:$ $P^{2}=P_{i}P^{i}$. Of course, in $11d$,
one can also obtain a lot of invariants if one considers those constructed
from $P_{i}$ and tensors $Z_{ij},Z_{ijklm}$. But it is not necessary, and
the minimal theory (i.e. $11d$ supergravity) can be constructed starting
from $P_{i}$ only. From the $SO(2,10)$ point of view we are forced to switch
on tensorial charges from the beginning. Here we shall consider the problem
of construction of field theories with tensor $P_{\mu \nu }$ instead of the
usual momenta $P_{i}$. The final goal is the construction of a theory with
$%
SO(2,10)$ invariance, which has to reduce to $11d$ theories under some
conditions. In particular, although the algebra $(\ref{2})$ has many
different BPS representations \cite{MM}, \cite{U}, there is one, which
corresponds to the usual BPS representation of $11d$ supergravity - massless
superparticle multiplet. This is the multiplet, corresponding to the orbit
of a particular tensor
\begin{eqnarray}
P_{\mu \nu } &=&(P_{0^{\prime }i},P_{ij}=0),  \label{4} \\
P^{2} &=&P_{0^{\prime }i}P^{0^{\prime }i}=0,  \label{5} \\
P^{0^{\prime }i} &=&(1,1,0,...,0)  \label{6}
\end{eqnarray}
>From the $SO(2,10)$ point of view, the (compact) little group of this orbit
is $SO(9)$, exactly as in $11d$ superalgebra, and the representations are
the same, in particular the smallest representation of the corresponding
superalgebra includes one spin-vector ($128$ degrees of freedom), a
traceless second-rank tensor $(44)$, and an antisymmetric third rank tensor
$%
(84)$. One should note also that the natural space on which the algebra with
$SO(2,10)$ and $P_{\mu \nu }$ generators is realized is not $12d$
space-time, but $66d$ ``space-time'' with coordinates $X^{\mu \nu }$ -
antisymmetric tensor of second rank. The orbit referred to has the property,
that all polynomial invariants $\left( \ref{3}\right) $ are equal to zero.
The entire little group of this orbit can be found from its Lie algebra:
Consider all elements of $so(2,10)$ algebra, which leave the tensor
$(\ref{4}%
)$ unchanged. Actually $(\ref{4})$ itself, with raised second index (we use
``mostly plus'' signature), is an element of $so(2,10)$, so we are seeking
its stabilizer in this algebra. It is easy to show, that the following
matrices of $so(2,10)$ are exactly all matrices, commuting with $(\ref{4})$:
\begin{equation}
\left( 
\begin{array}{ccccccc}
0 & a & a & 0 & 0 & ... & 0 \\ 
-a & 0 & 0 & b & c & ... & d \\ 
a & 0 & 0 & -b & -c & ... & -d \\ 
0 & b & b & 0 & e & ... & f \\ 
0 & c & c & -e & 0 & ... & g \\ 
... & ... & ... & ... & ... & ... & ... \\ 
0 & d & d & -f & -g & ... & 0
\end{array}
\right)  \label{m1}
\end{equation}

The algebra of matrixes $(\ref{m1})$ is a direct sum of the $so(1,1)$
algebra of matrices $(\ref{m1})$ with only non-zero entry $a$, and an
algebra, which is a semidirect sum of $so(9)$ (represented by matrixes (\ref
{m1}) with $a=b=c=...=d=0$) and an Abelian algebra of matrices $(\ref{m1})$
with non-zero elements $b,c,...,d$ only. The unitary finite-dimensional
representations of this little group algebra are those of $so(9)$
subalgebra, with other generators represented by zero, which is possible due
to the structure of the algebra.

The main feature of theories of this kind (i.e. those having $P_{\mu \nu }$
instead of $P_{i}$ in the algebra of symmetries) is the appearance of many
Lagrangians - even in the case of one scalar field - which are
simultaneously necessary for the description of the theory. Or,
equivalently, many equations of motion for the same field are necessary,
which is immediately connected with the existence of many invariants of $%
P_{\mu \nu }$.Their role i
s to bring the general function of the momenta $%
P_{\mu\nu}$ to a function on the orbit of $SO(2,10)$. The notion of a
symmetry generalizes in this case, as we shall see below, to the single
equation, which implies that a sum of variations of these actions, with
different infinitesimal variation of the same field in different actions, is
equal to zero.

In the following we shall discuss mainly the theories without supersymmetry
generators. Supersymmetry will appear in Section $6$. Also, an algebra of
symmetry will be generalized to the semidirect product of $so(2,q)$ and that
of $P_{\mu \nu }$ which are tensors under $SO(2,q)$ Lorenz rotation, and
components of $P_{\mu \nu }$ commute with themselves. We are mainly
interested in $q=10$, but some examples will be considered for $q=2.$ In
Sections $3,4,5$ scalar, spinor and vector fields, respectively, are
considered. Section $2$ contains the discussion of orbits in the case of
q=2. The conclusion is devoted to the discussion of results and prospects.

\section{Classification of orbits: Example of q=2}

The method of induced representations requires the classification of all
orbits of the group $SO(2,10)$ on the space of tensors $P_{\mu \nu }$. This
classification can be presented in a form of a list of representatives, one
for each orbit, socalled standard forms of $P_{\mu \nu }$. It is also
convenient to have an identification of orbits with values of invariants,
constructed from $P_{\mu \nu }$, i.e. (\ref{3}). This correspondence can be
easily obtained if standard forms are known. The classification for the $%
(2+10)d$ case is not known to us. It is known in a simpler case, e.g. for
Maxwell field strength in $(1+3)d$ (see, e.g. \cite{nov}). Here we shall
present this for the case of $SO(2,2)$. In this case the space of tensors $%
P_{\mu \nu }$ can be invariantly divided into two subspaces of self-dual and
antiself-dual tensors, which have, correspondingly, the forms (with second
index raised)
\begin{equation}
P_{\mu}^{+\nu}=\left( 
\begin{array}{llll}
\,0 & \,a & \,b & \,c \\ 
-a & \,0 & -c & \,b \\ 
\,b & -c & \,0 & -a \\ 
\,c & \,b & \,a & \,0
\end{array}
\right)  \label{sd}
\end{equation}
and 
\begin{equation}
P_{\mu }^{-\nu }=\left( 
\begin{array}{llll}
0 & k & l & n \\ 
-k & 0 & n & -l \\ 
l & n & 0 & k \\ 
n & -l & -k & 0
\end{array}
\right)  \label{asd}
\end{equation}
The triplets $(a,b,c)$ and $(k,l,n)$ transform under algebra $so(2,2)$ as
vectors with respect to the two $so(1,2)$ subalgebras of $%
so(2,2)=so(1,2)+so(1,2)$. Correspondingly (neglecting some subtleties
concerning factorization over discrete subgroups) these vectors can be
brought to the different forms depending on the values of their squares,
namely to the forms $(m,0,0),(0,0,m)$ and$(1,0,1)$. Substitution of these
forms into $(\ref{sd})$, $(\ref{asd})$ gives the standard forms of $P_{\mu
\nu }^{\pm }$. The invariants, which define to which orbit a given $P_{\mu
\nu }$ belongs are traces of squares of self-dual and antiself-dual parts of 
$P_{\mu \nu }$: 
\begin{equation}
P_{\mu }^{+\nu }P_{\nu }^{+\mu }=4(-a^{2}+b^{2}+c^{2})  \label{pp1}
\end{equation}
\[
P_{\mu }^{-\nu }P_{\nu }^{-\mu }=4(-k^{2}+l^{2}+n^{2})\label{pp2} 
\]
The cases of negative, positive, and zero values of these invariants
correspond to the three abovementioned forms of tensors $(\ref{sd})$ and $(%
\ref{asd})$. In particular, it is easy to see, that the case of both
invariants equal to zero gives the matrix of the form (\ref{4}), (\ref{5})
and (\ref{6}).

\section{Scalar field}

According to the little group method of construction of the unitary
representations of semidirect product groups, we have to choose a particular
value of $P_{\mu \nu }$, take a particular unitary representation of the
corresponding little group, which is, by definition, the subgroup of $%
SO(2,q) $, which leaves that particular $P_{\mu \nu }$ unchanged, and induce
this representation on the whole group. In field theories with the usual
Poincare symmetry it is known tha
t the same representations can be described
also in the language of fields and their equations of motion. In the
simplest case of scalar (trivial) representation of the little group it is
necessary to use the space of the usual functions on the orbit, which can be
described in this case as the space of functions $\Phi (P_{\mu \nu })$,
satisfying the equations of motion
\begin{eqnarray}
(TrP^{2}-2m_{1}^{2})\Phi (P_{\mu \nu }) &=&0,  \label{6a} \\
(TrP^{4}-2m_{2}^{4})\Phi (P_{\mu \nu }) &=&0,  \nonumber \\
&&...  \nonumber \\
(TrP^{12}-2m_{6}^{12})\Phi (P_{\mu \nu }) &=&0  \nonumber
\end{eqnarray}
The first is the equation of the usual Klein-Gordon type, the others are on
the same footing as the first one, and altogether they define functional $%
\Phi (P_{\mu \nu })$ on the orbit, which is characterized by the numbers $%
m_{1},m_{2},...,m_{6}$. It is easy to check that on the orbit $(\ref{4})$
this set of equations of motion reduces to the usual Klein-Gordon equation
in $11d$:
\begin{eqnarray}
(P_{0^{\prime }i}P^{0^{\prime }i} - m^{2}_{1})\Phi (P_{0i}) &=&0
\label{kg11} \\
m_{2}=m_{3}=m_{4}=m_{5}=m_{6} &=& m_{1}  \nonumber
\end{eqnarray}
Actions, giving $(\ref{6a})$, are, in a coordinate representations,
\begin{eqnarray}
S_{i}&=&\frac{1}{2}\int [dX^{\mu \nu }](\Phi (X)(\frac{\partial }{\partial
X^{\mu _{1}\mu _{2}}}...\frac{\partial }{\partial X^{\mu _{2i-1}\mu _{2i}}}%
-m_{i}^{2i})\Phi (X))  \label{ActionS} \\
i&=&1,2,...6.  \nonumber
\end{eqnarray}
The question, whether it is possible to replace this set of Lagrangians by a
single one, giving all necessary equations (\ref{6a}), is open, even in the
simplest case of the scalar field. For the other hand, field theory can be
considered as a second-quantized version of the theory of particles. This
particle Hamiltonian, giving the same equations (\ref{6a}) as an equation
for the wave function of the quantized particle, can be represented in a
standard form:
\begin{equation}
H=\lambda _{1}(TrP^{2}-2m_{1}^{2})+...+\lambda _{6}(TrP^{12}-2m_{6}^{12})
\label{H}
\end{equation}
where $\lambda _{i}$ are arbitrary gauge functions (consequently, functions,
multiplying them, are constraints to be imposed on a wave function, cf. Eqs.
(\ref{6a})). We can define the Lagrangian after exclusion of momenta in the
usual relation:
\begin{equation}
L=P_{\mu \nu }\dot{X}^{\mu \nu } - H  \label{L}
\end{equation}
It may be worth mentioning, that a very similar problem was discussed many
years ago by Fiertz and Pauli \cite{FP}\footnote{%
We are indebted to M. Vasiliev for bringing this article to our attention.}
They considered, particularly, the problem of construction of the Lagrangian
for higher spin massive theories. These theories are described by a few
equations of the same field, as in the present situation, and Fierz and
Pauli developed the method of introduction of auxiliary fields, leading to a
single Lagrangian, with an equivalent set of equations of motion.

\section{Dirac equation(s)}

We now consider the spinor $\Psi$ as a function of tensorial momenta $P_{\mu
\nu }$ with the following set of equations of motion:
\newpage
\begin{eqnarray}
\Gamma ^{\mu _{1}\mu _{2}}P_{\mu _{1}\mu _{2}}\Psi (P_{\mu \nu }) &=&0
\label{7.1} \\
\Gamma ^{\mu _{1}\mu _{2}\mu _{3}\mu _{4}}P_{\mu _{1}\mu _{2}}P_{\mu _{3}\mu
_{4}}\Psi (P_{\mu \nu }) &=&0  \label{7.2} \\
\Gamma ^{\mu _{1}\mu _{2}\mu _{3}\mu _{4}\mu _{5}\mu _{6}}P_{\mu _{1}\mu
_{2}}P_{\mu _{3}\mu _{4}}P_{\mu _{5}\mu _{6}}\Psi (P_{\mu \nu }) &=&0
\label{7.3} \\
\Gamma ^{ \mu _{1}\mu _{2}...\mu _{8}}P_{\mu _{1}\mu _{2}}P_{\mu _{3}\mu
_{4}}P_{\mu _{5}\mu _{6}}P_{\mu _{7}\mu _{8}}\Psi (P_{\mu \nu }) &=&0
\label{7.4} \\
\Gamma ^{\mu _{1}\mu _{2}...\mu _{10}}P_{\mu _{1}\mu _{2}}P_{\mu _{3}\mu
_{4}}...P_{\mu _{9}\mu _{10}}\Psi (P_{\mu \nu }) &=&0  \label{7.5} \\
\Gamma ^{\mu _{1}\mu _{2}...\mu _{12}}P_{\mu _{1}\mu _{2}}......P_{\mu
_{11}\mu _{12}}\Psi (P_{\mu \nu }) &=&0  \label{7.6}
\end{eqnarray}
and actions:
\begin{eqnarray}
S_{i}&=&\frac{1}{2}\int [dX^{\mu \nu }](\bar{\Psi}(X) (\Gamma ^{\mu _{1}\mu
_{2}...\mu _{2i-1}\mu _{2i}}\frac{\partial }{\partial X^{\mu _{1}\mu _{2}}}%
...\frac{\partial }{\partial X^{\mu _{2i-1}\mu _{2i}}})\Psi(X))
\label{ActionF} \\
i&=&1,2,...6, \,\,\,\,\,\,\,\, \bar{\Psi}=\Psi^{T}\Gamma^{0}\Gamma^{0^{%
\prime}}  \nonumber
\end{eqnarray}
Again we can check that on the orbit$(\ref{4})$ this set is equivalent to
the $11d$ Dirac equation:
\begin{equation}
\Gamma ^{0^{\prime }i}P_{0^{\prime }i}\Psi (P_{0^{\prime }i})=0
\label{dir11}
\end{equation}
Then one can check that on-shell scalar invariants are zero $(\ref{3}):$%
\begin{eqnarray}
TrP^{2}\Psi (P_{\mu \nu }) &=&0,  \label{8.1} \\
TrP^{4}\Psi (P_{\mu \nu }) &=&0,  \label{8.2} \\
&...&  \nonumber \\
TrP^{12}\Psi (P_{\mu \nu }) &=&0.  \label{8.6}
\end{eqnarray}
To prove this, one has to multiply equations (\ref{7.1}) to (\ref{7.6}) on
different matrices like:
\[
\Gamma ^{\mu \nu }P_{\mu \nu },\,\,\,\Gamma ^{\mu \nu \lambda \rho }P_{\mu
\nu }P_{\lambda \rho },\,\,....
\]
and make some $\Gamma $-algebra calculations. For example, squaring ($\ref
{7.1})$, we have:
\begin{eqnarray}
\Gamma ^{\mu _{1}\mu _{2}}P_{\mu _{1}\mu _{2}}\Gamma ^{\mu _{3}\mu
_{4}}P_{\mu _{3}\mu _{4}}\Psi (P_{\mu \nu })&=&  \nonumber \\
\left(\Gamma ^{\mu _{1}\mu _{2}\mu _{3}\mu _{4}}P_{\mu _{1}\mu _{2}}P_{\mu
_{3}\mu _{4}} + 2TrP^{2}\right)\Psi (P_{\mu \nu })&=&0  \label{pr1}
\end{eqnarray}
i.e. Eq.. (\ref{8.1}). Multiplying ($\ref{7.1})$ on $\Gamma ^{\mu _{1}\mu
_{2}}P_{\mu _{1}\mu _{2}}^{3}$ we obtain
\begin{eqnarray}
\Gamma ^{\mu _{1}\mu _{2}}P_{\mu _{1}\mu _{2}}^{3}\Gamma ^{\mu _{3}\mu
_{4}}P_{\mu _{3}\mu _{4}}\Psi (P_{\mu \nu })&=&  \nonumber \\
\left( \Gamma ^{\mu _{1}\mu _{2}\mu _{3}\mu _{4}}P_{\mu _{1}\mu
_{2}}^{3}P_{\mu _{3}\mu _{4}}+2TrP^{4}\right) \Psi (P_{\mu \nu })&=&0
\label{9.2}
\end{eqnarray}
Squaring ($\ref{7.2})$ and using (\ref{8.1}) and higher equations from
($\ref{7.1}$) to ($\ref{7.6}$), we obtain
\begin{eqnarray}
\left( \Gamma ^{\mu _{1}\mu _{2}\mu _{3}\mu _{4}}P_{\mu _{1}\mu _{2}}P_{\mu
_{3}\mu _{4}}\right) ^{2}\Psi (P_{\mu \nu }) &=&  \nonumber \\
\left( 32\Gamma ^{\mu _{1}\mu _{2}\mu _{3}\mu _{4}}P_{\mu _{1}\mu
_{2}}^{3}P_{\mu _{3}\mu _{4}}+16TrP^{4}\right) \Psi (P_{\mu \nu }) &=&0
\label{9.3}
\end{eqnarray}
From $\left( \ref{9.2}\right) $, $\left( \ref{9.3}\right) $ and
$\left( \ref{8.1}\right)$ follows $\left( \ref{8.2}\right) $ and so on. Equivalently,
it is possible to use instead of ($\ref{7.1}$) to ($\ref{7.6}$) equations
($\ref{7.1}$) and
\begin{eqnarray}
G_{i}(P)\Psi (P_{\mu \nu }) =0, \,\,\,\,\,\, i = 2,3,4,5,6  \label{15.2a}
\end{eqnarray}
Here $G_{i}$ are combinations of invariants defined below in $\left( \ref{15.3}\right)$. 
Equations $\left( \ref{15.2a}\right)$ express higher
invariants through $TrP^2$ and $\left( \ref{7.1}\right)$ gives the on-shell
condition $TrP^2=0$ and fixes the spinor structure of $\Psi$.

\section{Vector field and gauge invariance}

We consider the Abelian vector field satisfying the usual Maxwell equation:
\begin{equation}
P_{j}F^{ji}=\left( -P^{2}\eta ^{ij}+P^{i}P^{j}\right) A_{j}=0  \label{10}
\end{equation}
This equation has a U(1) gauge invariance:
\begin{equation}
\delta A_{i}=P_{i}\alpha (P_{i})  \label{11}
\end{equation}
which allows one to impose a gauge fixing condition off-shell and remove the
longitudinal component:
\begin{equation}
P^{i}A_{i}=0  \label{12}
\end{equation}
Assuming $P^{2}\neq 0$ we can rewrite the Maxwell equation (\ref{10}) as:
\begin{equation}
P^{j}P_{j}A_{i}^{tr}=P^{2}\left( \delta _{i}^{j}-\frac{P_{i}P^{j}}{P^{2}}%
\right) A_{j}=0  \label{13}
\end{equation}
This equation means that $A_{i}^{tr}=0$ which implies the absence of
physical degrees of freedom. Therefore on-shell we have to put $P^{2}=0$ and
$P^{i}A_{i}=0$ which means in an appropriate frame: $P_{i}=(1,1,0,0,...)$
and $A_{0}=A_{1}$. These relations are equivalent to well-known result that
an on-shell vector field has $d-2$ physical components with the massless
condition $P^{2}=0$.

We now turn to the construction of the vector field theory in an $SO(2,10)$
invariant way. Using the vector $A_{\mu }(P_{\lambda \nu })$ and the ``field
strength''
\begin{equation}
F_{\mu \nu \lambda }=P_{\mu \nu }A_{\lambda }+P_{\nu \lambda }A_{\mu
}+P_{\lambda \mu }A_{\nu }  \label{14}
\end{equation}
we can define the following set of equations of motion:
\begin{eqnarray}
P^{\mu \nu }F_{\mu \nu \lambda } &=&0  \label{15-0.1} \\
(P^{3})^{\mu \nu }F_{\mu \nu \lambda } &=&0  \label{15-0.2} \\
&&... \\
(P^{11})^{\mu \nu }F_{\mu \nu \lambda } &=&0  \label{15-0.4}
\end{eqnarray}
An equivalent set of equations is:
\begin{eqnarray}
P^{\mu \nu }F_{\mu \nu \lambda }
&=&2\left( P_{\lambda }^{2\mu }-\frac{1}{2}%
TrP^{2}\delta _{\lambda }^{\mu }\right) A_{\mu }=0  \label{15.1} \\
G_{i}(P)A_{\mu } &=&0\,\,\,\,\,\,\,i=2,3,4,5,6  \label{15.2}
\end{eqnarray}
Here $G_{i}$ with $G_{1}=-\frac{1}{2}TrP^{2}$ are combinations of invariants
$\left( \ref{3}\right) $ coinciding with the coefficients in the expansion
of the characteristic polynomial for matrix $P_{\mu }^{\quad \nu }$:
\begin{eqnarray}
f(x) &=&\det (P-x)  \label{15.3} \\
&=&x^{12}+x^{10}G_{1}+x^{8}G_{2}+...+G_{6}  \nonumber
\end{eqnarray}
Again this set is equivalent to the ordinary Maxwell system under the
condition $P_{ij}=0$ and therefore reproduces the masslessness condition $%
P^{2}=0.$

We can now define the set of actions corresponding to equations
$(\ref{15.1}%
) $ to $(\ref{15.2})$ in coordinate space:
\begin{eqnarray}
S_{1} &=&\int [dX^{\mu \nu }]\left( -\frac{1}{6}F^{\mu \nu \lambda }F_{\mu
\nu \lambda }\right) ,  \label{15.4} \\
S_{i} &=&k_{i}\int [dX^{\mu \nu }]\left( \frac{1}{2}A_{\lambda
}G_{i}(\partial _{X^{\mu \nu }})A^{\lambda }\right) ,  \label{15.5} \\
i &=&2,3,..6.  \nonumber
\end{eqnarray}
Here we introduced coupling constants $k_{i}$ with appropriate
dimensionality. Actions, corresponding to the other forms of equations of
motion (\ref{15-0.1}) to (\ref{15-0.4}) are:
\begin{eqnarray}
K_{1} &=&\int [dX^{\mu \nu }]\left( -\frac{1}{6}F^{\mu \nu \lambda }F_{\mu
\nu \lambda }\right) ,  \label{k1} \\
K_{i} &=&k_{i}\int [dX^{\mu \nu }]\left( \frac{1}{2}A^{\mu }(\delta _{\mu
\nu }Tr(P^{2i-2})-2(P^{2i-2})_{\mu \lambda })A^{\lambda }\right) , \\
i &=&1,2,...6.
\end{eqnarray}
Another possible set of actions is
\begin{eqnarray}
N_{1} &=&\int [dX^{\mu \nu }]\left( -\frac{1}{6}F^{\mu \nu \lambda }F_{\mu
\nu \lambda }\right) , \\
N_{i} &=&k_{i}\int [dX^{\mu \nu }]\left( -\frac{1}{6}F^{\mu \nu \lambda
}(P^{2i-2})_{\lambda }^{\sigma }F_{\mu \nu \sigma }\right) , \\
i &=&1,2,...6.
\end{eqnarray}
What about gauge invariance? Let's define the following variation of field
A$%
_{\mu }$:

\begin{equation}
\delta A_{\mu }(P_{\lambda \rho })=P_{\mu \nu }\alpha ^{\nu }\delta
(G_{2}(P))...\delta (G_{6}(P))  \label{d1}
\end{equation}

Eqs. of motion (\ref{15.2}) are evidently invariant with respect to this
transformation, (\ref{15.1}) also is invariant, because delta-functions in
(%
\ref{d1}) put $P_{\mu \nu }$ to the form (\ref{4}), and (\ref{15.1}) becomes
usual Maxwell equation with usual gauge transformation (\ref{d1}). It is
easy to see, that gauge transformation (\ref{d1}) permits one to gauge away
$%
A_{0^{^{\prime }}}$ component and longitudinal part of $A_{i}$ (taking into
account that $A_{\mu }$ is non-zero only on the shell of delta-functions $%
\delta (G_{2}(P))$, ...,$\delta (G_{6}(P))$):

\begin{eqnarray}
\delta A_{0^{\prime }}(P_{0^{\prime }i}) &=&P_{0^{\prime }\nu }\alpha ^{\nu
}(P_{0^{\prime }i})=\beta  \label{d2} \\
\delta A_{i}(P_{0^{\prime }i}) &=&P_{i\nu }\alpha ^{\nu }(P_{0^{\prime
}i})=P_{i0^{\prime }}\gamma  \label{d3}
\end{eqnarray}
So, these equations describe the vector representation of SO(9) group, as
desired. Note that actions (\ref{15.4}), (\ref{15.5}) are invariant also.

Another way of realizing the same idea of gauge invariance is as follows. We
can define the following set of variations corresponding to each equation of
the set:
\newpage
\begin{eqnarray}
\delta _{1}A_{\mu }(P_{\lambda \rho }) &=&P_{\mu \nu }^{10}\alpha ^{\nu
}(P_{\lambda \rho })  \nonumber \\
\delta _{2}A_{\mu }(P_{\lambda \rho }) &=&k_{2}^{-1}P_{\mu \nu }^{8}\alpha
^{\nu }(P_{\lambda \rho })  \label{16} \\
&.&  \nonumber \\
&.&  \nonumber \\
\delta _{6}A_{\mu }(P_{\lambda \rho }) &=&k_{6}^{-1}\alpha _{\mu
}(P_{\lambda \rho })  \nonumber
\end{eqnarray}
Then it is easy to see that the following equation is satisfied:
\begin{equation}
\delta _{1}S_{1}+\delta _{2}S_{2}+....+\delta _{6}S_{6}=0  \label{17}
\end{equation}
due to the well-known Hamilton-Cayley identity for characteristic
polynomials:
\begin{equation}
f(P)=0  \label{18}
\end{equation}
where
\begin{equation}
f(x)=\det (P-x)  \nonumber
\end{equation}
As above, assuming that higher eqns. $\delta S_{i}/A_{\mu }$ $(i=2,...,6)$
are satisfied, eqn. (\ref{17}) gives the usual statement of gauge invariance
of Maxwell's action. (The equation similar to (\ref{17}) is valid for other
sets of actions, with different variations $\delta _{i}A_{\mu }$.) Then,
under the condition $P_{ij}=0$ the remaining symmetry is:
\begin{eqnarray}
\delta A_{0^{\prime }}(P_{0^{\prime }i}) &=&P_{0^{\prime }\nu }^{10}\alpha
^{\nu }(P_{0^{\prime }i})=\beta   \label{19} \\
\delta A_{i}(P_{0^{\prime }i}) &=&P_{i\nu }^{10}\alpha ^{\nu }(P_{0^{\prime
}i})=P_{i0^{\prime }}\gamma   \label{20}
\end{eqnarray}
The first one can be used for gauging away the additional twelfth component
$%
A_{0^{\prime }}$, the second gives the usual gauge transformation for the
remaining $11d$ Abelian gauge field. Of course, this symmetry is the same as
(\ref{d3}), (\ref{d3}) above The subtlety in (\ref{19}), (\ref{20}) is that:
if we consider the on-shell condition $P^{2}=0$ from the beginning, it is
not possible to gauge away $A_{0^{\prime }},$ as is seen from (\ref{19}).

Eqn. (\ref{17}) in this form is appropriate for generalization to
non-quadratic Lagrangians, but we are not aware on any general
considerations proving that such an equations are direct consequence of a
multilagrangian nature of the theory.

\section{Supersymmetry}

In this section we will consider $(2+2)d$ supersymmetric theories realizing
the following algebra of supersymmetry:
\begin{eqnarray}
\left\{ \bar{Q},Q\right\} &=&\Gamma ^{\mu \nu }P_{\mu \nu }  \label{toy} \\
\mu \nu ,... &=&0^{\prime },0,1,2.  \nonumber
\end{eqnarray}
We first define actions and the supersymmetry transformation for the
supersymmetry multiplets in the $(2+2)d$ theory, containing Majorana spinor
and scalar and spinor and vector fields. For the scalar (Wess-Zumino)
multiplet we can define the actions :
\begin{eqnarray}
S_{1}&=&\frac{1}{2}\int [dX^{\mu \nu }]\left(\Phi(\frac{\partial }{\partial
X^{\mu _{1}\mu _{2}}}\frac{\partial }{\partial X_{\mu _{1}\mu _{2}}})\Phi +
\bar{\Psi}\Gamma^{\mu _{1}\mu _{2}}\frac{\partial }{\partial X^{\mu
_{1}\mu_{2}}}\Psi\right)  \label{ActionWZ1} \\
S_{2}&=&k_{2}\frac{1}{2}\int [dX^{\mu \nu }]\left(\Phi G_{2}\Phi +
\bar{\Psi}
\Gamma^{\mu _{1}\mu _{2}\mu _{3}\mu _{4}}\frac{\partial }{\partial X^{\mu
_{1}\mu _{2}}}\frac{\partial }{\partial X^{\mu _{3}\mu _{4}}}\Psi\right)
\label{ActionWZ2}
\end{eqnarray}
and variations of fields:
\begin{eqnarray}
\delta _{1}\Phi &=&\bar{\varepsilon}\Psi, \,\,\,\,\,\, \delta _{1}\Psi =%
\frac{1}{2}\Gamma^{\mu _{1}\mu _{2}}\frac{\partial }{\partial X^{\mu
_{1}\mu_{2}}}\Phi\varepsilon  \label{var1} \\
\delta _{2}\Psi &=&-k^{-1}_{2}\Phi\varepsilon ,\,\,\,\,\,\, \delta _{2}\Phi
= 0  \label{var2}
\end{eqnarray}
The commutator of transformation (\ref{var1}) is $\Gamma ^{\mu \nu }P_{\mu
\nu }$, according to algebra (\ref{toy}), the commutator of (\ref{var2}) is
zero.

It is easy to check the following supersymmetry relation:
\begin{equation}
\delta _{1}S_{1}+\delta _{2}S_{2}=0  \label{SWZ}
\end{equation}
Again, as in the case of gauge invariance, the statement of supersymmetry is
of the kind of (\ref{17}): the sum of variations of actions of different
dimensionality is equal to zero. And again we can consider only variation of
first action (\ref{ActionWZ1}) with respect of transformations (\ref{var1})
. This transformation is the symmetry of (\ref{ActionWZ1}) if our fermionic
field is satisfying the equation of motion following from second action
(\ref
{ActionWZ2}).

Similar equations for the vector multiplet are:
\begin{eqnarray}
S_{1} &=&\int [dX^{\mu \nu }]\left( -\frac{1}{6}F^{\mu \nu \lambda }F_{\mu
\nu \lambda } + \frac{1}{2}\bar{\Psi}\Gamma ^{\mu _{1}\mu _{2}} \frac{%
\partial}{\partial X_{\mu _{1}\mu _{2}}}\Psi\right) ,  \label{vectS1} \\
S_{2} &=&k_{2}\int [dX^{\mu \nu }]\left( \frac{1}{2}A_{\lambda
}G_{2}(\partial _{X^{\mu \nu }})A^{\lambda }\right.  \nonumber \\
&+& \left. \frac{1}{2}\bar{\Psi} \Gamma^{\mu _{1}\mu _{2}\mu _{3}\mu_{4}}%
\frac{\partial }{\partial X^{\mu _{1}\mu_{2}}} \frac{\partial }{\partial
X^{\mu _{3}\mu _{4}}}\Psi\right),  \label{vectS2} \\
\delta _{1}A_{\mu } &=&\bar{\varepsilon}\Gamma _{\mu }\Psi  \label{27} \\
\delta _{1}\Psi &=&F_{\mu \nu \lambda }\Gamma ^{\mu \nu \lambda }\varepsilon
\label{28} \\
\delta _{2}\Psi &=&-k^{-1}_{2}\frac{1}{2}A_{\mu }\Gamma ^{\mu }\varepsilon
\label{29} \\
\delta _{2}A_{\mu} &=& 0 \\
\delta _{1}S_{1}&+&\delta _{2}S_{2} = 0
\end{eqnarray}
It is possible to find additional symmetries in the above actions which,
however, may not be important since all these theories are free.

\section{Conclusion and Outlook}

In the above we considered the construction of field theories, the
space-time algebra of symmetries of which is a semidirect sum of $so(2,q)$
and the Abelian algebra of second-rank tensors $P_{\lambda \rho }.$ The
reason is that the bosonic part of the $M$-theory algebra has this form,
with $q=10$, modulo a sixth-rank central charge, and the aim of the present
investigation is the development of an $SO(2,10)$-invariant approach to $M$%
-theory. This algebra differs from the usual Poincare one in two respects:
the momentum $P_{i}$ is replaced by $P_{\mu \nu }$, and second time-like
direction appears in the metric. The free theories of scalar, spinor, vector
fields are constructed, as well as the supersymmetric free theory for $q=2$.
One of the main features, which is different in the present situation from
the usual ones, is the necessity of having simultaneously many Lagrangians
for the same field. This follows from the existence of many invariants,
constructed from $P_{\mu \nu }$, which all have to be fixed for a given
unitary representation of algebra of symmetry. For a scalar field, e.g.,
these equations are Klein-Gordon ones and their higher derivative analogs.
Another peculiarity is the form of the gauge and susy invariance. It is
possible to write out the gauge transformations (for vector field), which is
the symmetry of all  equations of motion. But exist another formalism, which
seems to be more appropriate for generalisations, when the Lagrangians
mentioned above are not separately invariant with respect to these
transformations, but only the sum of their variations - with different
variations of the same field in different Lagrangians - is equal to zero. It
is well-known, that free theories can possess very different kinds of
algebras of symmetries, and only interaction selects the very few, such as
susy algebra. So, it is important to have an example of an interacting
theory. This can be constructed for $q=2$. The similar ``example'' for
$q=10$
would be an explicitly $SO(2,10)$-invariant formalism for $11d$
supergravity, unknown so far. It is possible to construct that on a
linearized level. On this stage we can propose the first level equations of
motion for graviton $h_{\mu \nu }$, gravitino $\Psi _{\mu }$, and third rank
antisymmetric tensor $A_{\mu \nu \lambda }$:
\begin{eqnarray}
-\frac{1}{2}\partial _{\alpha \beta }\partial ^{\beta \alpha }h_{\mu \nu
}-\partial _{\mu }^{\,\,\,\,\alpha }\partial _{\nu }^{\,\,\,\,\beta
}h_{\alpha \beta }+\partial _{\mu }^{2\,\,\alpha }h_{\alpha \nu }+\partial
_{\nu }^{2\,\,\alpha }h_{\alpha \mu }-\partial _{\mu \nu }^{2}h &=&0 \\
\gamma ^{\mu \nu \lambda \rho }\partial _{\nu \lambda }\Psi _{\rho } &=&0 \\
\partial ^{\mu \nu }F_{\mu \nu \lambda \rho \sigma } = 0\,\,\,\,\,\,\,
F_{\mu\nu \lambda \rho \sigma } &=&\partial _{[\mu \nu }A_{\lambda \rho \sigma ]}
\label{sugra}
\end{eqnarray}
It is easy to see that these equations go to usual $11d$ linearized
supergravity equations on the orbit (\ref{4}). The main problem here is the
formulation of high level equations of motion with complete set of
symmetries (global SUSY and gauge), which will be discussed separately. The
other interesting question is the construction of extended objects with
explicit $SO(2,10)$ invariance, appropriate for this hypothetical form of $%
11d$ supergravity. We refer to \cite{Hef1},\cite{Tsey} and \cite{Rud},\cite
{Rud1} for some suggestions.

\section{Acknowledgements}

This work was supported in part by the U.S. Civilian Research and
Development Foundation under Award \#96-RP1-253 and by INTAS grant \#96-538
and \#99-590. R.Mkrtchyan is indebted to M. Vasiliev and A. Melikyan for discussion. R.
Manvelyan is indebted to O. Lechtenfeld and H.J.W. M\"uller-Kirsten for
discussion and to the A. von Humboldt Foundation for financial support.

\end{document}